\documentclass[a4paper,11pt]{article}
\usepackage[left=3cm,right=3cm,
top=3cm,bottom=3cm]{geometry}
\usepackage[T2A]{fontenc}
\usepackage[utf8]{inputenc}
\usepackage{authblk}
\usepackage[english]{babel}
\usepackage{amsmath,amssymb,amsthm,mathtools}
\usepackage{euscript,mathrsfs}
\usepackage{icomma}
\usepackage{graphicx}
\usepackage{wrapfig}
\usepackage[dvipsnames]{xcolor}
\usepackage{indentfirst}
\usepackage{amsbsy}
\usepackage{wasysym}
\usepackage{cancel}
\usepackage{verbatim}
\usepackage{empheq}
\usepackage{adjustbox}
\usepackage{cite}
\usepackage[pdfencoding=auto]{hyperref}
\definecolor{urlcolor}{HTML}{120099}
\definecolor{linkcolor}{HTML}{005F5F}
\hypersetup{pdfstartview=FitH,  linkcolor=linkcolor,urlcolor=urlcolor, colorlinks=true,citecolor=blue}

\usepackage[framemethod=tikz]{mdframed}
\newmdenv[innerlinewidth=0.5pt, roundcorner=4pt,linecolor=blue,innerleftmargin=6pt,
innerrightmargin=6pt,innertopmargin=6pt,innerbottommargin=6pt]{mybox}

\renewcommand{\phi}{\varphi}

\DeclareMathOperator{\Tr}{Tr\,}

\begin{document}

\title{\bf Page time and the order parameter for a consciousness state}
\author[a,b]{Alexander Gorsky}

%\affil[a]{Moscow Institute of Physics and Technology, Dolgoprudny 141700, Russia}
\affil[a]{Institute for Information Transmission Problems, Moscow 127994, Russia} 
\affil[b]
{Laboratory of Complex Networks, Brain and Consciousness Research Center, Moscow}

\maketitle

\centerline{\bf Abstract}
In this Letter using the analogy with the recent resolution of the 
black hole information paradox we conjecture 
the order parameter for the state of consciousness based on the notion of
the Page curve and the Page time. 
The entanglement between the state of the brain and time series 
of neuronal firing as well as the non-orthogonality of the functional
connectomes play a key role.

\section{Introduction and motivation}

The  origin of the 
 consciousness certainly is the  interesting scientific problem.
 It is desirable to identify the guidance
 principle behind the phenomenon and there are several
 ideas formulated to handle with this issue. 
 In the integrated information theory(IIT)
 \cite{tononi1998consciousness,tononi2004information} it is assumed that the entanglement of the
 clusters in the times series data for the neuronal firing could play the role
 of the order parameter of the consciousness state. The identification of the clusters
 can be simplified via the spectral analysis of the 
 network formed from the time series \cite{toker2019information}.
 The idea of the global neuronal workspace  is based on the
 version of the self-organized criticality \cite{dehaene2011experimental,mashour2020conscious}.
 It is assumed that interaction between the neuronal groups in nonlinear 
 and transition to the consciousness state is a kind of phase transition.
 Another combination of ideas yields the model of competing cellular
assemblies \cite{crick2003framework}. The neuronal group selection theory 
focuses at the evolutionary aspects of the formation of consciousness \cite{edelman2004wider}.

More physically motivated 
 thermodynamical approach via  a kind of the
 free energy principle was suggested in \cite{friston2010free}.
 Some ideas based on the quantum decoherence were discussed
 in \cite{hameroff2014consciousness}. 
The review of the measures of consciousness can be found in \cite{seth2006theories} while the recent review of the most popular theories of consciousness can be found in \cite{anokhin2021cognitome}.

   Several features are quite common for all models 
   \begin{itemize}
       \item The state of  consciousness is related to  consolidation of  effective degrees of freedom  in the brain - functional connectomes
       \item There is  specific timescale attributed to the formation of  consciousness state from the non-conscious state
       \item Some version of entanglement serves as the order parameter for the  consciousness state
       \item There is the interplay between the deterministic component  and the ensemble average in the  formation of  consciousness state
       \item The cluster structure in the time series is important in some models
       \item The evolution of the system is unitary
       
   \end{itemize}
We shall argue that these features can be naturally unified  within the framework of the Page curve and Page time \cite{page1,page2} assuming the entanglement between the brain states and  time series of neuron firing.

The Page curve has been invented to keep unitarity  in a two-component 
system in a pure random state taking into account properly the entanglement 
of subsystems.
The entanglement is evaluated via the density matrix
projected into the subsystem \cite{page2}. The spectrum of 
projected density matrix before the Page time obeys the Marchenko-Pastur (MP) law
but gets modified after the Page time.
This approach has been applied for the entanglement of a black hole (BH) and its radiation \cite{page1} however 
the BH example is not unique and in principle one can consider any other "black box" entangled with its own radiation or time series of some activity.

The BH information paradox has been known since 70-ths \cite{hawking1976breakdown} and it claims that
the entropy production of the Hawking
BH radiation violates a unitarity at large times, see \cite{almheiri2021entropy,raju2022lessons} for the recent reviews. The resolution of the information paradox has been found
very recently and the island conjecture has been formulated.
It was argued  \cite{penington2020entanglement,almheiri2019entropy} that the state of BH radiation is encoded in some island subregions behind the BH horizon. This has been quantified in the simplified model via end of world branes(EOW)
\cite{kourkoulou2017pure}, each of them has large number of internal degrees of freedom
\cite{Almheiri_2020,penington2022replica}. The number of EOW branes grows with time and BH with EOW branes get entangled  with the BH radiation.
This is the first ingredient important for the resolution of the information paradox.

The second crucial ingredient goes as follows.
The EOW branes interact  and can  develop the wormhole non-perturbative configurations 
connecting the pairs of the EOW branes. At the Page time the EOW branes get consolidated into the single disc geometry since  the multiple non-perturbative wormholes start to dominate \cite{penington2022replica}. After the Page time the entropy of radiation goes down and the unitarity gets restored. It is important that after the Page time the Petz map known in the information theory \cite{petz1986sufficient,wilde2013quantum} allows to map 
the operations inside the BH Hilbert space into the operations in the Hilbert space of BH radiation \cite{penington2022replica}. Such mapping is impossible before the Page time. If the deterministic component is
introduced into the system \cite{Blommaert_2022,blommaert2022microstructure} a kind of perturbative interaction between the EOW branes is induced.

We conjecture that somewhat similar picture emerges statistically during   formation of  consciousness state
from the non-conscious one. The Hilbert space of brain and the Hilbert space of  time series together are considered as the two-component entangled system. We assume that the functional connectomes are analogue of the EOW branes and encode the information concerning the time series of neuronal activity. The functional connectomes are non-orthogonal and interact in two ways. First they interact as multilayer networks and secondly there is the "contact interaction" since some neurons belong to several functional connectomes simultaneously.
The state of brain with ensemble of
functional networks is entangled with the time series  and the number of functional networks involved grows in time. At   time of formation of a consciousness state the functional networks
get consolidated similar to the consolidation of EOW branes and the entropy attributed to time series stops to grow. 
 
 Hence we assume that the Page curve describes the time evolution of the entropy production in the time series 
 of neuronal firing
and at the Page time the functional connectomes get consolidated which is necessary condition for the
establishing of a consciousness state. We conjecture that the sign of time derivative
of the  entropy production in the time series and clustering of the networks  formed from the time series can serve as the indicators for a 
 state of consciousness.

\section{BH toy model}
\subsection{The entanglement of the BH and radiation}
 Let us recall the notion of the Page curve for a 
 entangled two-component system. 
It was shown in \cite{page2} that the entanglement 
spectrum of the system  in pure random state involving
two subsystems and projected to one of them  provides the 
important information concerning unitarity.
Let rectangular matrix $X$ describes uniformly random pure quantum state in $C^n\otimes C^m$. 
Then $Y=XX^+$- reduced density matrix
on $C^n$ whose eigenvalue distribution depends on the evolution time.
At small times  spectral density of entanglement matrix $Y$ obeys MP distribution
\cite{page2}. However at some time, called the Page time the spectrum of the 
entanglement matrix gets modified and the entropy production in the subsystem 
which we have projected on starts to decrease. The entropy behaves 
according to Page curve as follows
\begin{equation}
  S(n,m) = \left(\log n -\frac{n}{2m}\right)\theta(m-n) +\left(\log m -\frac{m}{2n}\right)\theta(n-m)
\end{equation}
If the number of degrees of freedom in one subsystem grows with time the 
change of regime occurs exactly at the Page time, 

The Page curve has been recognized recently 
in the context of the resolution of the BH information paradox. In this case 
the entangled two-component system is as follows. The one subsystem involves 
$e^S_{BH}$ degrees of freedom where $S_{BH}= \frac{A}{4G_N}$  is the Bekenstein-Hawking BH entropy
and $A$ is the area of horizon.
The second system describes the ongoing radiation of the BH and  dimension of
the corresponding Hilbert space
$k$ grows in time. These two subsystems are entangled and 
it  was argued \cite{penington2020entanglement} that there is 
quantum extremal surface Q behind the BH horizon \cite{engelhardt2015quantum}
 encoding the BH radiation. The island conjecture assumes
 that the entropy of BH radiation reads as 
 \begin{equation}
     S(\rho_R)= \min\left[\frac{Area(Q)}{4G_N} + S_{\text{bulk}}\right]
 \end{equation}
 Here $\rho_R$ is the density matrix of the BH radiation  in the full theory
coupled to quantum gravity. The second term corresponds to the 
von Neumann entropy of the region in the interior of the BH bounded 
by the quantum extremal surface Q. At small times the second contribution is small
however at large times the leading saddle in the  
semi-classical path integral on the Euclidean black hole 
gets modified and the different extremal surface starts to
dominate. The key point is that in this new phase the 
entropy of the radiation decreases in agreement with a unitarity.
The corresponding behaviour with the Page time is presented in Fig.1

\begin{figure}
    \centering
    \includegraphics[width=0.55\linewidth]{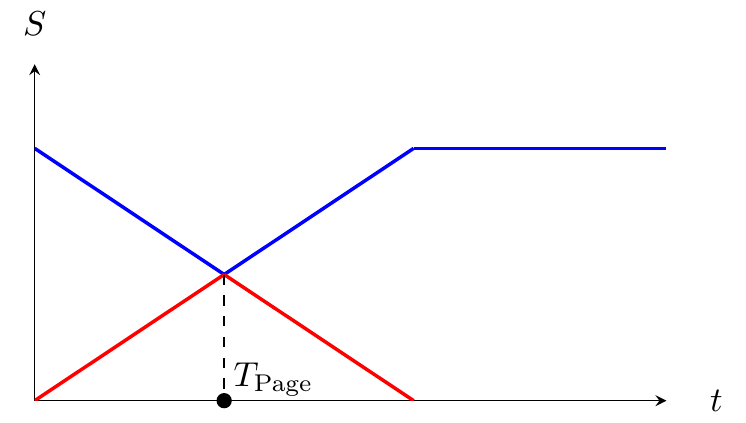}
    \caption{Page curve for the entropy of BH radiation(red)}
    \label{fig:picture}
\end{figure}

%\begin{figure}
    %\centering
%    \includegraphics[width=\linewidth]{picture.pdf}
%    \caption{Page curve for the entropy of BH radiation(red)}
%    \label{fig:picture}
%\end{figure}

 In the simplified version instead of the extremal surface one considers
 multiple EOW branes \cite{penington2022replica,Almheiri_2020} and 
 analyses what is the saddle point configuration  at
 the different times.
 Naively one considers the states
 \begin{equation}
      \sum_{i=1}^k \left|\Psi_i\right\rangle\otimes\left|i\right\rangle
 \end{equation}
   where $\Psi_i$ is the state of BH with the non-dynamical EOW branes behind the horizon 
 while the state $i$ belongs to the auxiliary basis of ongoing Hawking modes \cite{penington2022replica}.  
 The states $\Psi_i$ are not orthogonal  which is the second important point  behind the explanation
 of the BH information paradox
 \begin{equation}
     \left\langle\Psi_i|\Psi_j\right\rangle = e^{2S_{BH}}\delta_{ij} +O\left(e^{S_{BH}}\right).
 \end{equation}
 The second term corresponds to the contribution from non-perturbative wormholes or effect of
 more direct EOW brane interactions and generically is subject to the ensemble averaging with 
 some measure.
 Quantitatively the density matrix projected onto the radiation reads as 
 \begin{equation}
     \rho_R=\sum_{i,j}^k\left\langle\Psi_j|\Psi_i\right\rangle|i\rangle\langle j|= \sum_{i,j}^k\sum_{a=1}^{e^S}C^{*}_{aj}C_{ai}|i\rangle\langle j|
 \end{equation} 
 In the simplest case the random rectangular $e^{S_{BH}}\times k $ matrices $C$ are averaged with the Gauss
 measure with unit covariance
 \begin{equation}
     \int dC dC^{+} \exp(-\Tr C^{+}C)
 \end{equation}
 The spectral density of $\rho_R$ has been evaluated in \cite{penington2022replica}
 via solution to the Schwinger-Dyson equation for the resolvent  in the planar 
 approximation which takes into account the replica wormholes between the 
 EOW branes.
 In microcanonical ensemble with unit Gaussian distribution it reads as
\begin{equation}
\rho_R(x)= \frac{ke^S_{BH}}{2\pi x}\sqrt{[x-(k^{-1/2}- e^{S_{BH}/2})^2][(k^{-1/2}+ e^{S_{BH}/2})^2-x)} +
\delta(x)(k-e^S_{BH})\theta(k-e^S_{BH})
\end{equation}
We see that its smooth part coincides with the  MP distribution.

 At some value of k the entropy of the BH and the entropy evaluated from the 
 radiation density matrix $\rho_R$
 coincide. This happens at the Page time and the
 interpretation from the 
 viewpoint of internal degrees of freedom behind the horizon
 has been developed in \cite{penington2022replica}. It has been
 argued that  the semiclassical saddle point for EOW branes before the Page time
implies that they can be considered as the separated degrees of freedom which non-perturbatively
 interact via wormhole solutions. However upon the summation of the replica
 wormholes at large times another saddle point dominates and the EOW branes form the single disk-like geometry
 which should be considered as single consolidated state of multiple EOW branes see Fig 2. 
 The off-diagonal terms in the matrix elements start to dominate. After
 the Page time the Petz map \cite{petz1986sufficient,wilde2013quantum} allows to map the operations in the
 radiation Hilbert space into the operations in the  BH Hilbert space \cite{penington2022replica}.

\begin{figure}
    \centering
    \includegraphics{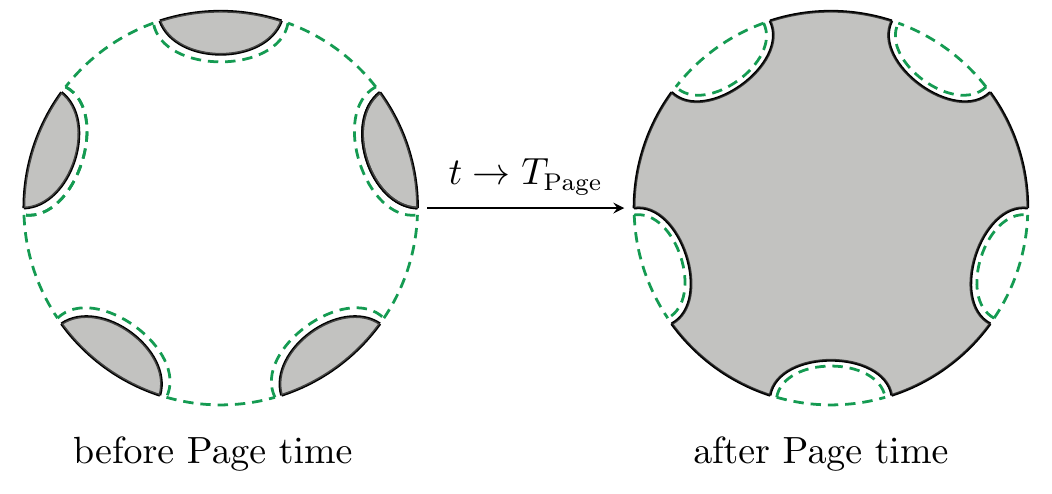}
    \caption{Islands behind the horizon are entangled with BH radiation. At the Page time
    the multiple islands get consolidated into the disc-like geometry.}
    \label{fig:Islands_Page}
\end{figure}

\subsection{BBP transition and entanglement spectrum }
We have argued above that at the Page time the entanglement spectrum
of the radiation gets modified and some number of the eigenvalues
get separated from continuum after the Page time. This follows from the effect of non-orthogonality 
of the internal states of the BH. 
Recently it has been suggested  \cite{blommaert2022microstructure} that the admixture of the determinism
in the averaging over ensemble can induce the "perturbative" interaction
of the EOW branes. Moreover it properly induces the eigenvalue separation
phenomenon in the MP distribution necessary to get the Page curve.

The phenomenon of the eigenvalue separation is familiar in the probability theory
 when we perturb the Gaussian measure for random variables. We are interested 
 in the deformation of the measure involved into the projected density matrix
 in the pure random states that is in the Wishart ensemble. Two types of measure
 deformation are known; the shifted Gaussian measure $\exp(-\Tr(H-H_0)^2)$ when the
 expansion around the nontrivial saddle takes place \cite{forrester} and the 
 measure with non-null covariance $\exp(-\Tr X^+X\Sigma)$ \cite{baik2005phase,baik2006painleve}.
 In both cases there is the critical value of parameters involved in $H_0$ or $\Sigma$
 when one or several eigenvalues get decoupled from the bulk of the spectrum. It is
 a kind of phase transition (BBP transition) with own universality class and the 
 distribution of the largest eigenvalue obeys the so-called spiked Tracy-Widom(TW) law exactly at the
 transition point.

To fit with our conjecture we modify the Gaussian measure not by the shift of 
mean value as in \cite{blommaert2022microstructure} but by introducing non-null complex covariance which 
does the same job \cite{baik2005phase,baik2006painleve}. 
This goes as follows. Let us consider $M$ vectors $(\vec{y_1}\dots \vec{y_M})$ each of dimension $N$. The density of all vectors is Gaussian with  covariance $\Sigma$, where $\Sigma$ is positive $N\times N$ matrix 
\begin{equation}
P(\vec{y})= \frac{1}{(2\pi)^{N/2}(\det\Sigma)^{1/2}}
\exp\left(-\frac{1}{2}\left\langle\vec{y},\Sigma^{-1},
\vec{y}\right\rangle\right)
\end{equation}

Consider the $N\times M$ matrix $X=\left[{\bf y}_1 - {\bf Y}, \dots {\bf y}_M - {\bf Y} \right]$ and combine it into  $N\times N$ covariance matrix $S=XX^{t}$. Let us assume that $M\to \infty$ and $N\to \infty$ such that their ratio is finite $\frac{M}{N}=\gamma$. The matrix $S$ belongs to the Laguerre or Wishart ensemble whose asymptotic properties at $\Sigma=Id$ are known. The spectral density obeys the MP law
\begin{equation}
\rho(x)= \frac{\gamma^2}{2x}\sqrt{(b-x)(x-a)}, \qquad a<x<b 
\end{equation}
where
\begin{equation}
a=\left(\frac{\gamma-1}{\gamma}\right)^2;\qquad b=\left(\frac{\gamma+1}{\gamma}\right)^2
\end{equation}
The largest eigenvalue fixing the spectral edge is $\lambda_{max}=b$ and the spectral fluctuations at the spectral edge obey the TW distribution
\begin{equation}
P\left((\lambda_{max} - b) \frac{\gamma}{(1+\gamma)^{4/3}} M^{2/3}\leq s\right) \rightarrow  F_{GOE}(s)
\end{equation}
where $F_{GOE}$ is the TW distribution for orthogonal ensemble. Now departure from the identity matrix 
is introduced and some number of degenerate non-unit eigenvalues of the covariance matrix $\Sigma$ are 
selected $l_1 \dots l_k\neq 1$. The value $l_1$ is considered as the control parameter for the BBP
phase transition.

For complex Gaussian samples \cite{baik2005phase} the critical value of the non-unit covariance eigenvalue for perturbed case has been found $l_{crit}= 1 +\gamma^{-1}$ where $\gamma$ is a parameter of the ensemble.
\begin{equation}
P\left((\lambda_{max}-l_{crit}^2) \frac{\gamma}{(1+\gamma)^{4/3}} N^{2/3}\leq s\right) \rightarrow  F_k(s)
\end{equation}
The function $F_k(s)$ is the limiting distribution, where the largest eigenvalue $l_1$ of the covariance matrix has the multiplicity $k$ and is equal to the critical value, hence we have single perturbation parameter. If $k$ non-unit eigenvalues of the covariance matrix are above the critical value $l_{crit}$, i.e. $l_i> l_{crit}$, distribution of  $\lambda_{max}$ obeys the distribution in $k\times k$ Gaussian unitary ensemble, $G_k(s)$. 
\begin{equation}
P\left((\lambda_{max} - \left(l_1 + \frac{l_1 \gamma^{-2}}{l_1-1}\right)\right) N^{1/2}\sqrt{l_1^2 -  \frac{l_1^2 \gamma^{-2}}{(l_1-1)^2}} \leq s) \rightarrow  G_k(s)
\end{equation}

The function $F_k(s)$ can be generalized to the function $F_k(s,w_1,\dots,w_k)$ if all perturbation parameters are different. Similarly to the pure TW case, the distribution obeys some solution to the Painleve II equation with particular monodromy properties \cite{baik2006painleve}. In general case the determinantal representation of $F_k(s,w_1, \dots,w_k)$ is available in terms of the single function of two variables $f(s,w)$ which satisfy the pair of differential equations depending on the Painleve II solutions -- see \cite{baik2006painleve} for details.

Summarizing, there are three regimes for the spectral density
depending on the relation between $k$ and $e^{S_{BH}}$. 
The transition at the  Page time indicates
that the new saddle in the gravitational path integral
corresponding to
connected geometry with multiple  wormholes starts to dominate at $k\propto e^{S_{BH}}$. 
In terms of the eigenvalues
exactly at the Page transition $k-e^{S_{BH}}$ eigenvalues get escaped from 
the bulk of distribution and settle at $\lambda=0$. 
If somebody would be interested in the question if all EOW branes get consolidated
or not one should look at the sign of the time derivative for the radiation 
entropy and at the number of isolated eigenvalues. This is the order parameter separating the phases.

\section{ Towards the  order parameter for a consciousness state}

\subsection{Entropy for the time series}
Since we will assume that the time series of the brain activity 
is the part of the whole two-component system it is necessary 
to provide the tool to evaluate  the entropy of the time series 
similarly to the entropy of the BH radiation. The effective way
goes as follows, first we represent the time series data via the 
network and than evaluate the  entropy of this network.

There are at least three ways to generate the network
from the time series (see, \cite{varley2021network} for the review).
The first way involves the introduction of the effective parameter which 
quantifies the vicinity in the data set and the nodes within
this effective radius are connected by links.
In the second approach the notion of visability is introduced and 
the horizontally visable nodes are connected by links. Finally,
in the third approach a kind of Markov dynamics for the probability
is formulated and its Markov operator yields the network
adjacency matrix.

Once we find the network representation for the time series some
standard entropy measured can be applied. The simplest one is
the conventional thermodynamic entropy defined through the 
density metrics for the network
\begin{equation}
\rho= \frac{\exp(-\beta L)}{Z}
\end{equation}
where $L$ is the graph Laplacian or normalized graph Laplacian,
and $\beta$ is auxiliary parameter which allows to scan the spectrum.
The thermodynamical entropy is defined as 
\begin{equation}
    S= \Tr\rho \log \rho
\end{equation}
and is useful to determine the characteristic time scales 
for the diffusion in the network  \cite{de2016spectral}.

However we are interested in the entropy production hence we have 
to consider the coarse-grained entropy instead of the fine-grained entropy
which is constant in time. The useful coarse-grained information entropy production measure has
been discussed in \cite{lynn2021broken} which is suitable for our purpose,
\begin{equation}
\dot{S}=\sum_{ij} P_{ij} log (\frac{P_{ij}}{P_{ji}}) 
\end{equation}
where  $P_{ij}=Prob[x_{t-1}=i,x_t=j]$ and $x_t$ corresponds to the state of the system at time $t$.
It is nothing but the Kullback-Leibler divergence between the forward and reverse transition probabilities.
This measure for the entropy production for different tasks has been investigated 
experimentally  \cite{lynn2021broken}. It was shown that the entropy production
vanishes or is very small in non-conscious state, is quite substantial for the cognitive task and 
maximal at the physical activity. Remark that the entropy production provides
the thermodynamical time arrow and it was noted recently that in the interacting
many-body system the time arrow emerges via the interaction only \cite{lynn2022decomposing}.

\subsection{The conjecture}

To formulate our conjecture 
remind that the brain involves structural and
multiple functional networks. The structural network just 
reflects the connections between neurons and is completely
known for C'Elegance with 302 nodes only. Each functional
network or functional connectome involves all neurons specialized in some function.
They are connected to each other and can be located in
the very different parts of the brain. The functional networks
are assumed to interact and the individual neurons can
belong to the several functional networks simultaneously.
It is generally believed that in the state of consciousness
the functional connectomes get consolidated via some mechanism.

Let us apply the Page's ideas to the brain and assume that
the functional connectomes play the similar role as EOW
branes in the BH example while the spiking time series
play the role of BH radiation. Hence we have the "black box"
with some degrees of freedom inside and its radiation and would
like to investigate the phase of internal state of the brane 
looking at the properties of  time series only. 

Let us follow the same logic as before and assume that 
the state of the system involves the state $\Xi_i$
which corresponds to the state of the brain with the
functional connectomes in the state i and the 
states $|i>$ which are the eigenfunctions of the Laplacian
of the graph representing the time series. Hence 
we consider the states 
\begin{equation}
\sum_{i} |\Xi_i\rangle\langle i|
\end{equation}
The density matrix projected
to the time series state $ \rho_{TS}$ is
\begin{equation}
 \rho_{TS}=\sum_{i,j}^k\langle\Xi_j|\Xi_i\rangle|i\rangle\langle j|= \sum_{i,j}^k\sum_{a=1}^{S_{brain}}C^{*}_{aj}C_{ai}|i\rangle\langle j|
 \end{equation}
 where $S_{brane}$ is the entropy of the brain in the non-conscious state. The value 
 of $k$ measures the number of functional connectomes which are switched on.
 
The projected density matrix yields
the entanglement entropy of the spiking time series $S_{TS}$.
To keep the unitarity we assume the 
existence of the Page curve and Page time when the 
internal functional networks get consolidated, see Fig.3, 
providing the state of   consciousness at the
Page time. After the Page time presumably the Petz map
yields the mapping of the operations within the time series 
to the operations between the functional connectomes in the brain.

\begin{figure}
    \centering
    \includegraphics{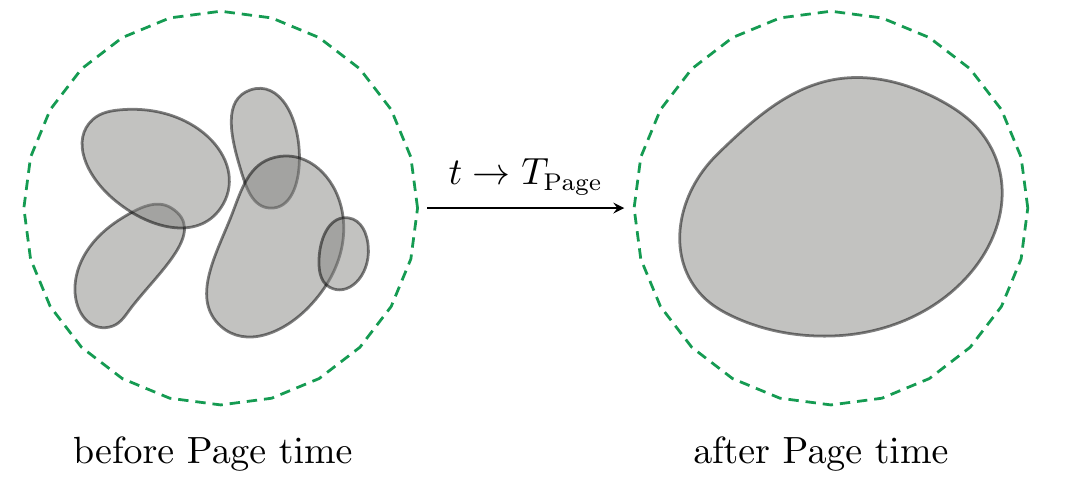}
    \caption{The functional connectome represents the network of neurons 
    with the same specialization. There are multiple functional connectomes
    in the brain which can overlap due to the multispecialized neurons. We 
    conjecture that  functional connectomes are entangled with the 
    time series of neuron activity and at the effective Page time when the
    state of consciousness gets established they get consolidated. }
    \label{fig:Connectomes_Page}
\end{figure}

As in the BH toy example the functional connectomes
overlap  and are non-orthogonal. The interaction of the
functional connectomes can have  perturbative and
non-perturbative contributions. The simplest one is
generated by " contact interaction" we have described 
above. In the case of a brain this interaction can be 
attributed to the neurons with several specializations.
Due to the overlap of functional connectomes  the  effective 
brain entropy and the entropy of the time series become of the same
order and the "common space " of the functional connectomes gets formed at the Page time. 
Note that the contact interaction  amounts to the non-unit covariance matrix
which induces the isolated eigenvalues in the spectrum 
corresponding to the clusters in the time series network \cite{nadakuditi2013spectra}.
Such clusters in the time series networks are the essential ingredients
of IIT.

Our conjecture fits with the general expected features of the 
consciousness state
\begin{itemize}
    \item The consolidation of the functional networks occurs at the Page time
    \item There is the timescale when consciousness state gets established
    \item Order parameter indeed is related to some specific entanglement
    \item Deterministic component is important. It induces the effective interactions
    and simultaneously generates the clusters in the time series

\end{itemize}
Summarizing the arguments above we  suggest that the sign of the time derivative
of  the entropy production in the time series  serves as the order parameter $R$
for a consciousness state
\begin{equation}
R=\mathrm{sign}\left(\frac {dS_{TS}}{dt}\right).
\end{equation}
Another important indicator is the increasing of the clusters in the
network formed from the time series.
We can not exclude that the entropy production stop raising at the Page
and decreases very slowly at the late times.

Hence it is natural to question how the predictions following from our 
conjecture can be measured
experimentally and if there are any contradictions with the existing experimental data. 
The simplest possibility concerns 
the investigation of the entropy production and clusterization 
of the time series during the  waking up process and the exit from the state of anesthesia. 
Hopefully some relevant results concerning the anesthesia state are available.
It was found in \cite{wenzel2019reduced}    that at exit from anesthesia two effects are observed
both for mice and humans (see also \cite{luppi2021brain,boyce2023cortical,bharioke2022general,nilsen2020proposed}).  First, the number of clusters in the networks formed by time series
(microstates in notations of \cite{wenzel2019reduced}) increases substantially  and secondly, consolidation of the functional
connectomes gets increased as well. Moreover both features are reversible while
depth of consolidation depends on the type of anesthesia applied \cite{varley2020differential}. This is consistent
with our conjecture.

Note also that the establishing time of 
consciousness state is different for the waking up and
the anesthesia exit processes. That is we can assume that the 
number of functional networks switched off at anesthesia
is larger then at sleep state. It would be very interesting to measure 
the entropy production at the anesthesia process to compare with our conjecture.

\section{Conclusion}

In this Letter we have conjectured using very general statistical arguments
for the unitary two-component entangled many-body system the new order parameter
for the consciousness state which measures if the functional connectomes
are consolidated or not. Certainly it is necessary but insufficient criteria. 
Two points are crucial for this conjecture.
First,it is the entanglement between 
the time series of the neuron activity and the brain state
involving multiple overlapping functional connectomes.
Secondly, the non-orthogonality of the states of functional connectomes
which is small at early time but starts to dominate at the effective  Page time.
The mechanism behind the resolution of the information paradox for a BH radiation 
provides the proper framework to analyze this problem.  The clustering of the
time series density matrix occurs naturally within this approach and is another
indication of the state of consciousness.

The conjecture is a bit speculative but we believe that 
this framework could be useful for this research area and
suggest the very clear-cut check. We do not
provide the microscopic  explanation of the consciousness state
restricting ourselves by the order parameter and do not
present the  hypothetical functional whose extremization  presumably yields 
two different saddle points. 
However we expect that 
some notions of the information geometry could be useful
in this respect. In particular the recent observation \cite{alexandrov2022information}
concerning the relation between the singularity of the 
information metric in the simplest all-to-all disordered Kuramoto 
model and the synchronization phase transition provides the example
of such relation. Indeed microscopically the consolidation
of the functional connectomes implies the synchronization phenomena
in the multiplex networks. Another approach concerning the quantum
information geometry approach can be found in \cite{georgiev2020quantum}. 
Note also that some analogy between the
brain and the BH in a  different setting has been mentioned 
in \cite{dvali2018black}. It would be interesting to combine the ideas
from that paper with our study.

I am grateful to A. Alexandrov, K. Anokhin, A. Milekhin, S. Nechaev and
N. Pospelov for the useful discussions and comments.
I want to thank IHES and Nordita where the parts of the work have been done for the hospitality and support. The work was supported by  grant \textnumero 075-15-2020-801 by Ministry of Science and Higher Education of Russian Federation.

%\newpage 
\bibliographystyle{unsrt}
\bibliography{references}
\end{document}